\documentclass{article}
\usepackage{amsfonts}
\usepackage{amsmath}
\usepackage{amsthm}
\usepackage{bm}
\usepackage[sorting=none,maxbibnames=5,giveninits]{biblatex}
\usepackage{fixltx2e}
\usepackage{float}
\usepackage[T1]{fontenc}
\usepackage[bottom]{footmisc}
\usepackage[margin=1in]{geometry}
\usepackage{graphicx}
\usepackage[utf8]{inputenc}
\usepackage{parskip}
\usepackage[section]{placeins}
\usepackage{subcaption}
\usepackage{url}
\usepackage{svg}

\DeclareNameAlias{author}{last-first}
\DeclareFieldFormat[article,incollection,inproceedings,techreport]{title}{#1} 

\newcommand{\D}{\mathcal{D}}
\newcommand{\N}{\mathcal{N}}

\DeclareMathOperator{\E}{\mathbb{E}}
\DeclareMathOperator{\bernoulli}{Bernoulli}

\makeatletter
\AtEveryCitekey{%
  \ifcsundef{blx@entry@refsegment@\the\c@refsection @\thefield{entrykey}}
    {\csnumgdef{blx@entry@refsegment@\the\c@refsection @\thefield{entrykey}}{\the\c@refsegment}}
    {}}
\defbibcheck{onlynew}{%
  \ifnumless{0\csuse{blx@entry@refsegment@\the\c@refsection @\thefield{entrykey}}}{\the\c@refsegment}
    {\skipentry}
    {}}
\makeatother

\title{Differences between human and machine perception in medical diagnosis}

\author{
Taro Makino$^{1,2,*}$,
Stanisław Jastrzębski$^{2,3,1}$,
Witold Oleszkiewicz$^{7}$,
Celin Chacko$^{2}$,\\
Robin Ehrenpreis$^{2}$,
Naziya Samreen$^{2}$,
Chloe Chhor$^{2}$,
Eric Kim$^{2}$,
Jiyon Lee$^{2}$,\\
Kristine Pysarenko$^{2}$,
Beatriu Reig$^{2,6}$
Hildegard Toth$^{2,6}$,
Divya Awal$^{2}$,
Linda Du$^{2}$,\\
Alice Kim$^{2}$,
James Park$^{2}$,
Daniel K. Sodickson$^{2,3,5,6}$,
Laura Heacock$^{2,6}$,
Linda Moy$^{2,3,5,6}$,\\
Kyunghyun Cho$^{1,4}$,
Krzysztof J. Geras$^{2,3,5,1,*}$\\\\

$^1$Center for Data Science, New York University\\
$^2$Department of Radiology, NYU Langone Health\\
$^3$Center for Advanced Imaging Innovation and Research, NYU Langone Health\\
$^4$Department of Computer Science, Courant Institute, New York University\\
$^5$Vilcek Institute of Graduate Biomedical Sciences, NYU Grossman School of Medicine\\
$^6$Perlmutter Cancer Center, NYU Langone Health\\
$^7$Faculty of Electronics and Information Technology, Warsaw University of Technology\\
$^*$\texttt{taro@nyu.edu, k.j.geras@nyu.edu}\\
}

\date{}

\begin{document}
\maketitle
\newrefsegment

\begin{abstract}
Deep neural networks (DNNs) show promise in image-based medical diagnosis, but cannot be fully trusted since their performance can be severely degraded by dataset shifts to which human perception remains invariant. If we can better understand the differences between human and machine perception, we can potentially characterize and mitigate this effect. We therefore propose a framework for comparing human and machine perception in medical diagnosis. The two are compared with respect to their sensitivity to the removal of clinically meaningful information, and to the regions of an image deemed most suspicious. Drawing inspiration from the natural image domain, we frame both comparisons in terms of perturbation robustness. The novelty of our framework is that separate analyses are performed for subgroups with clinically meaningful differences. We argue that this is necessary in order to avert Simpson's paradox and draw correct conclusions. We demonstrate our framework with a case study in breast cancer screening, and reveal significant differences between radiologists and DNNs. We compare the two with respect to their robustness to Gaussian low-pass filtering, performing a subgroup analysis for microcalcifications and soft tissue lesions. For microcalcifications, DNNs use a separate set of high frequency components than radiologists, some of which lie outside the image regions considered most suspicious by radiologists. These features run the risk of being spurious, but if not, could represent potential new biomarkers. For soft tissue lesions, the divergence between radiologists and DNNs is even starker, with DNNs relying heavily on spurious high frequency components ignored by radiologists. Importantly, this deviation in soft tissue lesions was only observable through subgroup analysis, which highlights the importance of incorporating medical domain knowledge into our comparison framework.
\end{abstract}

Following their success in the natural image domain~\cite{Krizhevsky2012AlexNet,Simonyan2015VGGNet,Ren2015FasterRCNN,RedmonYOLO2016,He2016ResNet,Huang2017DenseNet,He2017MaskRCNN}, deep neural networks (DNNs) have achieved human-level performance in various tasks of image-based medical diagnosis~\cite{Esteva2017SkinCancer,Lindsey2018FractureDetection,Coudray2018LungCancer,Liu2019AIDiagnosisReview,Wu2019Mammography,Shen2019GMIC,Shen2020GMIC,Rodriguez2019AI101Radiologists,Ardila2019LungCancer,McKinney2020GoogleMammography,Kim2020Lunit,Schaffter2020CombinedAIRadiologist,Liu2020SkinDiseases}. DNNs have a number of additional benefits: they can reach diagnoses quickly, do not suffer from fatigue, and can be deployed anywhere in the world. However, they currently possess an Achilles' heel which severely limits their clinical applicability. They cannot be fully trusted, given their extreme vulnerability to dataset shifts to which human perception is robustly invariant. For example, a dermatologist-level skin cancer classifier, approved for use as a medical device in Europe, learned to spuriously associate surgical skin markings with malignant melanoma~\cite{Winkler2019SurgicalSkinMarkings}. As a result, the classifier's false positive rate increased by 40\% during an external validation. Also, in three out of five cases, a pneumonia classifier performed substantially worse in cross-institutional settings than in the institution in which it was trained~\cite{Zech2018CrossHospitalGeneralization}.

A promising path to establishing trust in DNNs for medical diagnosis is to understand whether and how their perception is different to that of humans. With this understanding in hand, we can recognize the limits of DNNs and better anticipate their failures. An effective empirical means of comparing human and machine perception is to study their robustness to input perturbation. By removing certain information from the input and analyzing the resulting change in prediction, we can infer the degree to which that information was utilized. This approach has been successfully applied in the natural image domain, and has exposed fundamental differences between human and machine perception~\cite{Szegedy2014IntriguingProperties,Jo2017SurfaceStats,Dodge2017DistortionComparison,Geirhos2018HumansDNNs,Hendrycks2019RobustnessBenchmark,Yin2019Fourier}. We extend this line of work, taking into account a critically important consideration for medical diagnosis.

\begin{figure}
    \centering
    \includegraphics[width=\textwidth]{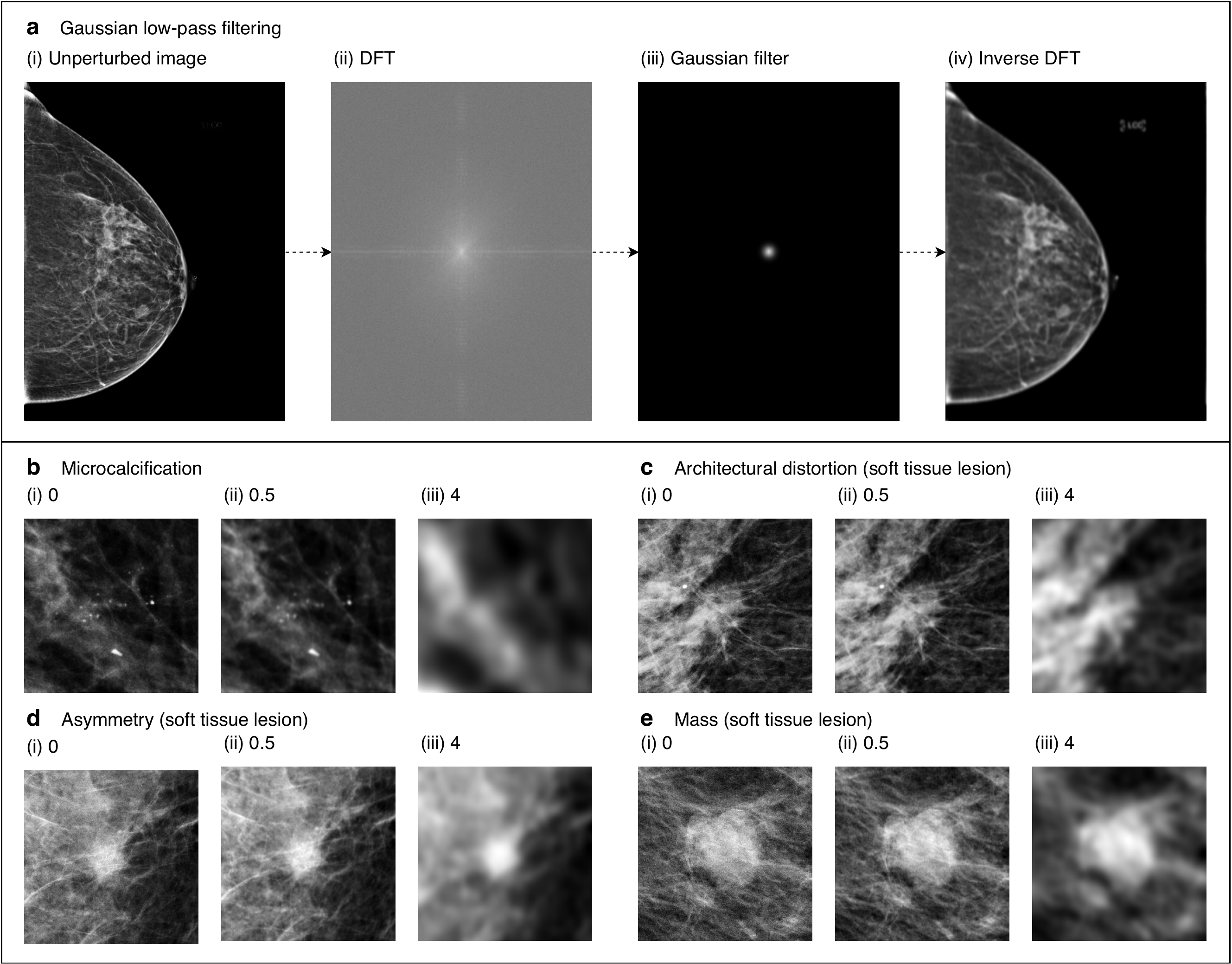}
    \caption{\textbf{Identification of subgroups and an input perturbation.} In our breast cancer screening case study, we separately analyzed two subgroups: microcalcifications and soft tissue lesions, using Gaussian low-pass filtering as the input perturbation. \textbf{(a)} Gaussian low-pass filtering is composed of three operations. The unperturbed image is transformed to the frequency domain via the two-dimensional discrete Fourier transform (DFT). A Gaussian filter is applied, attenuating high frequencies. The image is then transformed back to the spatial domain with the inverse DFT. \textbf{(b--e)} Gaussian low-pass filtering applied to various types of malignant breast lesions. Subfigures (i--iii) show the effects of low-pass filtering of increasing severity. \textbf{(b)} Microcalcifications are tiny calcium deposits in breast tissue that appear as white specks. Radiologists must often zoom in significantly in order to see these features clearly. Since these microcalcifications have a strong high frequency component, their visibility is severely degraded by low-pass filtering. \textbf{(c)} Architectural distortions indicate a tethering or indentation in the breast parenchyma. One of their identifying features are radiating thin straight lines, which become difficult to see after filtering. \textbf{(d)} Asymmetries are unilateral fibroglandular densities that do not meet the criteria for a mass. Low-pass filtering blurs their borders, making them blend into the background. \textbf{(e)} Masses are areas of dense breast tissue. Like asymmetries, masses generally become less visible after low-pass filtering, since their borders become less distinct. In our subgroup analysis, we aggregated architectural distortions, asymmetries, and masses into a single subgroup called ``soft tissue lesions.'' This grouping was designed to distinguish between localized and nonlocalized lesions. Soft tissue lesions on the whole are far less localized than microcalcifications, and they require radiologists to consider larger regions of the image during the process of diagnosis.}
    \label{fig:low-pass_filtering}
\end{figure}

\begin{figure}
    \centering
    \includegraphics[width=\textwidth]{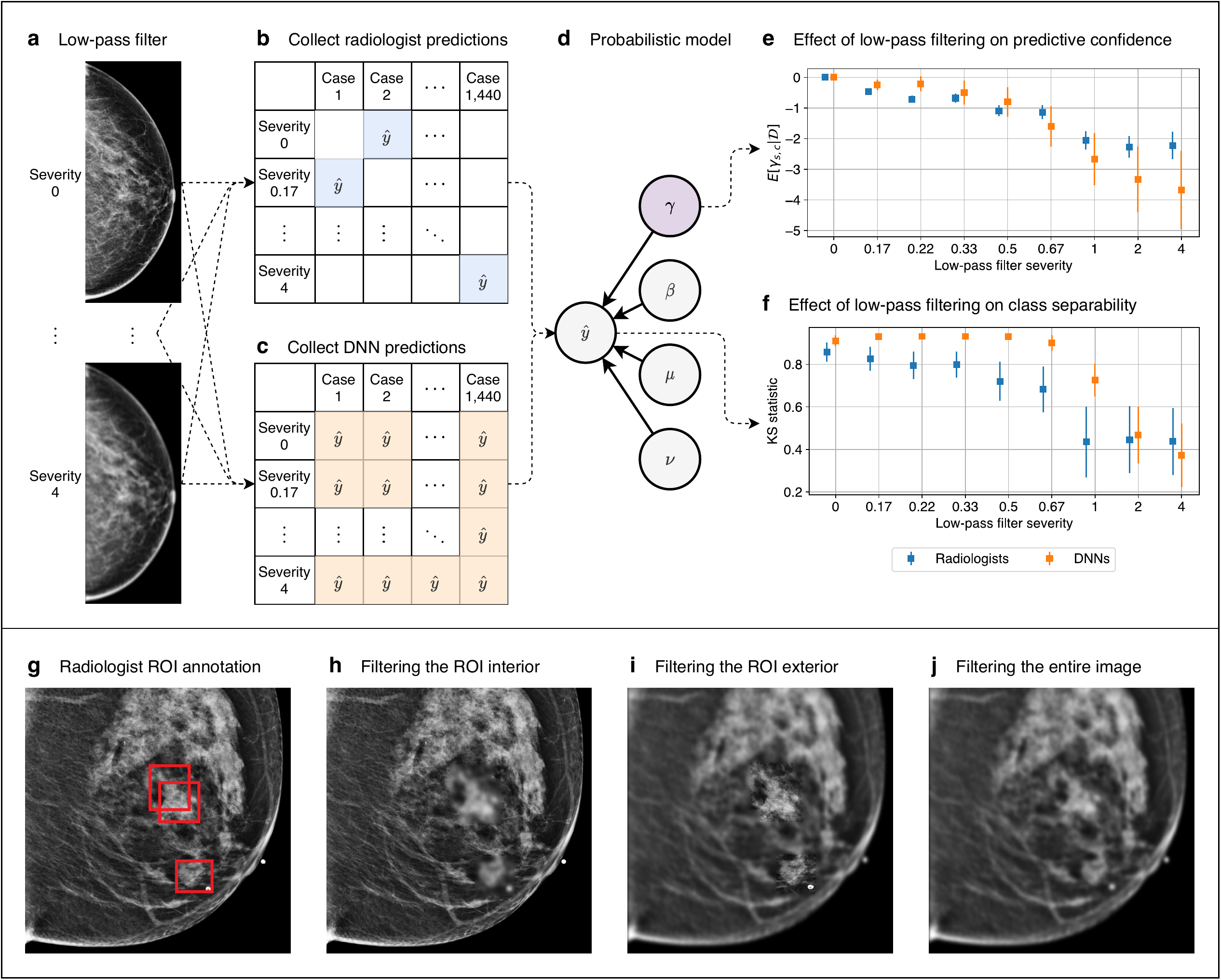}
    \caption{\textbf{Our framework applied to breast cancer screening.} \textbf{(a--f)} Comparison of radiologists and DNNs with respect to their sensitivity to the removal of clinically meaningful information. \textbf{(a)} We applied low-pass filtering to a set of mammograms using a wide range of filter severities. \textbf{(b)} We conducted a reader study in which each reader was provided with the same set of mammograms. Each reader saw each exam once, and each exam was filtered with a random severity. Thus, each radiologist's predictions populate a sparse matrix. \textbf{(c)} Predictions were collected from DNNs on the same set of exams. Unlike radiologists, DNNs made a prediction for all pairs of filter severities and cases, so their predictions form a dense matrix. \textbf{(d)} Probabilistic modeling was applied to the predictions, where a latent variable $\gamma$ measures the effect of low-pass filtering, and a separate variable $\eta$ factors out individual idiosyncrasies. \textbf{(e)} We examined the posterior expectation of $\gamma$ to evaluate the effect of low-pass filtering on predictive confidence. \textbf{(f)} We sampled from the posterior predictive distribution and computed the distance between the distributions of predictions for malignant and nonmalignant cases. This represents the effect that low-pass filtering has on class separability. \textbf{(g--j)} Comparison of radiologists and DNNs with respect to the regions of an image they find most suspicious. \textbf{(g)} Radiologists annotated up to three regions of interest (ROIs) that they found most suspicious. We then applied low-pass filtering to: \textbf{(h)} the ROI interior, \textbf{(i)} the ROI exterior, and \textbf{(j)} the entire image. We analyzed the robustness of DNNs to these three filtering schemes in order to understand the degree to which the DNNs utilize information in the interiors and exteriors of the ROIs.}
    \label{fig:framework}
\end{figure}

We argue that subgroup analysis is necessary in order to draw precise and correct conclusions regarding perception in medical diagnosis. From a performance standpoint, subgroups are important in medical diagnosis because different types of errors can vary greatly in clinical significance~\cite{OakdenRayner2020HiddenStratification}. When models are trained within the dominant paradigm of empirical risk minimization, all errors are treated equally, which can lead to a large disparity in performance across subgroups. Robust optimization addresses this issue by considering the performance of various subgroups, and optimizing the worst-case subgroup performance~\cite{Sagawa2019DRO,Goel2020ModelPatching}. In addition to performance, subgroups are also important for understanding perception. A failure to incorporate subgroups can lead to incorrect conclusions due to Simpson's paradox~\cite{Pearl2014SimpsonsParadox}, which states that subgroup-specific relationships can disappear or even reverse when the subgroups are aggregated. We therefore propose a framework which uses subgroup-specific perturbation robustness to compare human and machine perception in medical diagnosis. While others have separately studied perturbation robustness and subgroup performance, we combine the two ideas to draw precise conclusions regarding perception. We demonstrate our framework with a case study in breast cancer screening, and show that failure to account for subgroups would indeed result in incorrect conclusions. It is important to note that while we analyze perturbation robustness, our purpose here is not specifically to improve the robustness of machine-derived diagnosis. Rather, we aim to use perturbation robustness as a lens to understand human and machine perception.

In our framework, human and machine perception are compared along two axes, represented by two distinct questions. First, \emph{if we subject humans and machines to a perturbation which removes clinically meaningful information, do they share the same robustness characteristics?} Second, \emph{to what degree do humans and machines agree on the most suspicious regions of an image?}

The first step in this framework is to identify subgroups that are diagnosed in a significantly different manner, and to find an input perturbation that removes clinically relevant information from each of these subgroups. See Figure~\ref{fig:low-pass_filtering} for an illustration of this step applied to breast cancer screening. With the subgroups and perturbation identified, humans and machines can then be compared with respect to their perturbation robustness. Predictions are collected from humans and machines based on medical images to which the perturbation has been applied with varying severity. Probabilistic modeling is applied to these predictions, capturing the isolated effect of the perturbation, while factoring out confounding effects such as individual idiosyncrasies. The probabilistic model is used to compare the perturbation robustness of humans and machines in terms of two criteria that are important for diagnosis: predictive confidence and class separability. Predictive confidence measures the strength of predictions, and is independent of correctness. Class separability represents correctness, and is quantified as the distance between the distributions of predictions for positive and negative cases. See Figure~\ref{fig:framework} (a--e) for a visualization of this procedure for comparing human and machine predictions in the setting of breast cancer screening.

Humans and machines are then compared with respect to the regions of an image they find most suspicious. This is also framed in terms of perturbation robustness. First, humans annotate the most suspicious regions of interest (ROIs) for each image. The perturbation is then applied in separate steps to the interior of these ROIs, to the exterior regions only, and to the entire image. By analyzing the robustness of machines to these perturbations, we can infer whether humans and machines find the same regions informative. Figure~\ref{fig:framework} (g--j) illustrates this process for mammographic images.

Our framework enables a nuanced comparison of human and machine perception for medical diagnosis through the analysis of subgroup-specific perturbation robustness. In our case study, we examined the robustness of radiologists and DNNs to Gaussian low-pass filtering, separately analyzing two subgroups called microcalcifications and soft tissue lesions. We discovered significant differences between radiologists and DNNs in both subgroups. For microcalcifications, DNNs use a different set of high frequency components than radiologists, and some of these components lie outside image regions identified as most suspicious by radiologists. It is unclear from the current analysis whether these divergent frequency components represent spurious features or potential new biomarkers of disease. For soft tissue lesions, the difference in perception is even more significant, with DNNs using clearly spurious high frequency components ignored by radiologists. Importantly, we show that without subgroup analysis, we would have failed to observe this divergent behavior in soft tissue lesions, thus artificially inflating the similarity of radiologists and DNNs.

\section*{Results}

\paragraph{Experimental setup.}

\begin{figure}[h]
    \centering
    \includegraphics[width=\textwidth]{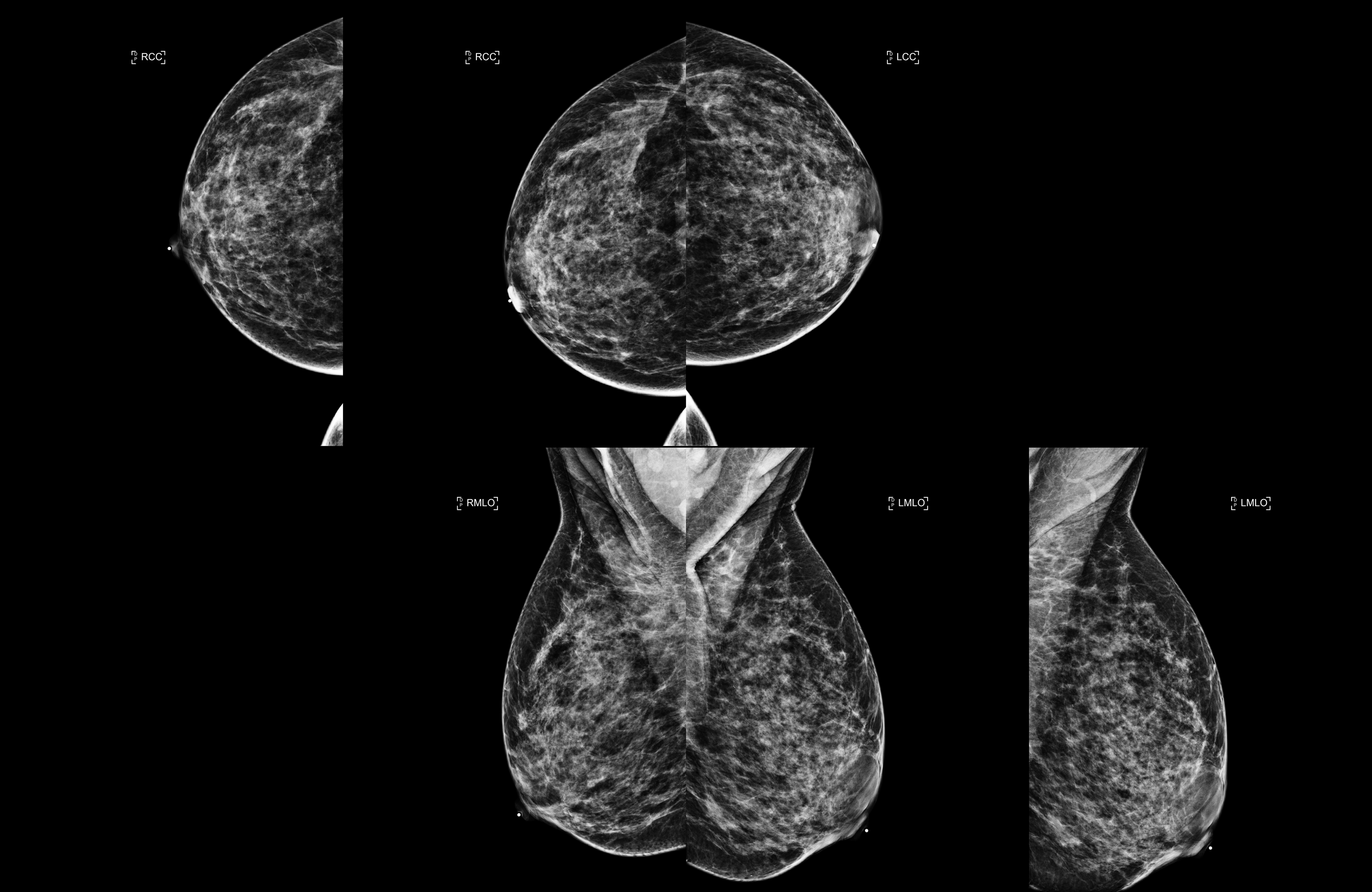}
    \caption{\textbf{An example of a screening mammogram.} An exam is composed of at least one image from each of four views. The views, in clockwise order, are: left craniocaudal (L-CC), left mediolateral oblique (L-MLO), right mediolateral oblique (R-MLO), and right craniocaudal (R-CC). This example shows that there can be multiple images per view. We present the images to radiologists in a format called ventral hanging, where the right breast faces left and is presented on the left, and the left breast faces right and is displayed on the right. Additionally, the craniocaudal (CC) views are on the top row, while the mediolateral oblique (MLO) views are on the bottom row. In contrast to radiologists, DNNs made predictions on each image individually after they were cropped to a consistent size.}
    \label{fig:mammogram}
\end{figure}

We experimented with the NYU Breast Cancer Screening Dataset~\cite{Wu2019NYUDataset} developed by our research team and used in a number of prior studies \cite{Wu2019Mammography,Shen2019GMIC,Shen2020GMIC,Fevry2019MammogramLocalization,Wu2020MultipleViews}, and we applied the same training, validation, and test set data split as has been reported previously. This dataset consists of 229,426 screening mammography exams from 141,473 patients. Each exam contains at least four images, with one or more images for each of the four standard views of screening mammography: left craniocaudal (L-CC), left mediolateral oblique (L-MLO), right craniocaudal (R-CC), and right mediolateral oblique (R-MLO). Each exam is paired with labels indicating whether there is a malignant or benign finding in each breast. See Extended Data Figure~\ref{fig:mammogram} for an example of a screening mammogram. We used a subset of the test set for our reader study, which is also the same subset used in the reader study of~\cite{Wu2019Mammography}. For our DNN experiments, we used two architectures in order to draw general conclusions: the deep multi-view classifier (DMV)~\cite{Wu2019Mammography}, and the globally-aware multiple instance classifier (GMIC)~\cite{Shen2019GMIC,Shen2020GMIC}. We primarily report results for GMIC, since it is the more recent and better-performing model. The corresponding results for DMV, which support the generality of our findings, are provided in the Extended Data section.

\paragraph{Perturbation reader study.}

In order to compare the perception of radiologists and DNNs, we applied Gaussian low-pass filtering to mammograms, and analyzed the resulting effect on their predictions. We selected nine filter severities ranging from unperturbed to severe, where severity was represented as a wavelength in units of millimeters on the physical breast. Details regarding the calculation of the filter severity are provided in the Methods section. Figure~\ref{fig:low-pass_filtering} demonstrates how low-pass filtering affects the appearance of malignant breast lesions.

We conducted a reader study in order to collect predictions for low-pass filtered images from radiologists. This reader study was designed to be identical to that of~\cite{Wu2019Mammography}, except that the mammograms were randomly low-pass filtered in our case. We assigned the same set of 720 exams to ten radiologists with varying levels of experience. The images were presented to the radiologists in a conventional format, and an example is shown in Extended Data Figure~\ref{fig:mammogram}. Each radiologist read each exam once, and for each exam, we uniformly sampled one severity level out of our set of nine, and applied it to all images in the exam. The radiologists made binary predictions indicating the presence of a malignant lesion in each breast. We describe the details of the reader study in the Methods section.

We then trained five DNNs from random weight initializations, and made predictions on the same set of 720 exams. We repeated this nine times, where the set of exams was low-pass filtered with each of the nine filter severities. We note that for each DNN, we made a prediction for every pair of exam and filter severity. In contrast, for each radiologist, we only had predictions for a subset of the possible combinations of exam and filter severity. This means that if we arrange the predictions in a matrix where each row represents a filter severity and each column an exam, the matrix of predictions is sparse for each radiologist, and dense for each DNN. This fact is visualized in Figure~\ref{fig:framework} (b--c). The sparsity of the radiologist predictions is by design; we were careful to ensure that each radiologist only read each exam once, since if they were to have seen the same exam perturbed with multiple filter severities, their predictions would have been unlikely to be independent. However, the sparsity prevents us from comparing radiologists and DNNs using evaluation metrics that use predictions for the complete set of exams. We therefore utilized probabilistic modeling to use the available radiologist predictions to infer values for the missing predictions.

\paragraph{A probabilistic model of predictions.}

\begin{figure}
    \centering
    \includegraphics[width=0.7\textwidth]{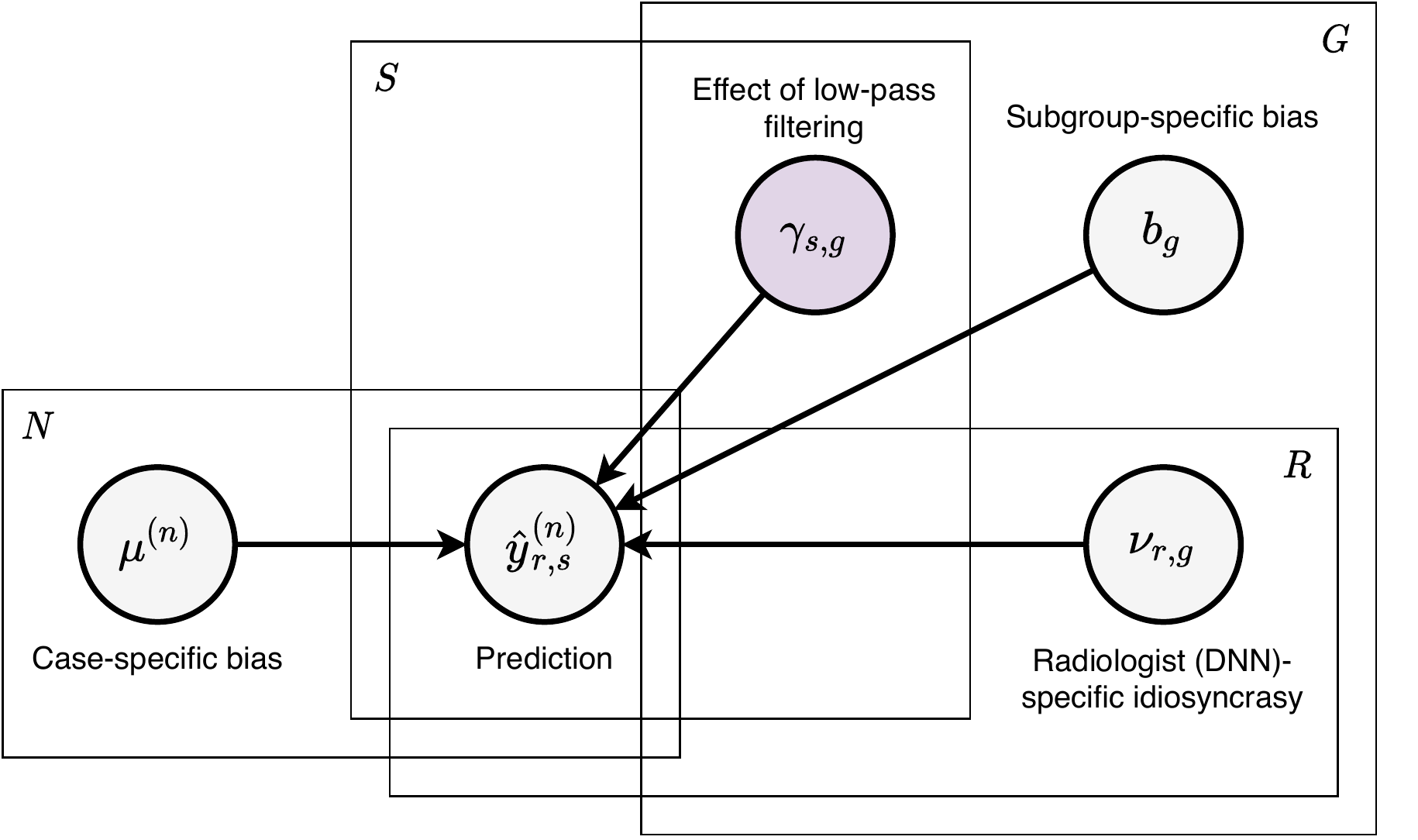}
    \caption{\textbf{Probabilistic model.} Our modeling assumption is that each prediction of radiologists and DNNs is influenced by four latent variables. $\hat{y}_{r, s}^{(n)}$ is radiologist (DNN) $r$'s prediction on case $n$ filtered with severity $s$. As for the latent variables, $b_g$ represents the bias for subgroup $g$, $\mu^{(n)}$ is the bias for case $n$, $\gamma_{s, g}$ is the effect that low-pass filtering with severity $s$ has on lesions in subgroup $g$, and $\nu_{r, g}$ is the idiosyncrasy of radiologist (DNN) $r$ on lesions in subgroup $g$. Our analysis relies on the posterior distribution of $\gamma_{s, g}$, as well as the posterior predictive distribution of $\hat{y}_{r, s}^{(n)}$. The other latent variables factor out potential confounding effects.}
    \label{fig:pgm}
\end{figure}

We applied probabilistic modeling to achieve two purposes. The first is to study the effect of low-pass filtering on specific subgroups of lesions in isolation, after factoring out various confounding effects such as the idiosyncracies of individual radiologists and DNNs. The second is to infer the radiologists' predictions for each pair of exam and filter severity, since some pairs were missing by design. We modeled the radiologists' and DNNs' predictions as i.i.d. Bernoulli random variables. Let us denote radiologist (DNN) $r$'s prediction on case $n$ filtered with severity $s$ as $\hat{y}_{r, s}^{(n)}$. We parameterized our model as
\begin{equation}
    \hat{y}_{r, s}^{(n)} \sim \bernoulli(\sigma(b_g + \mu^{(n)} + \gamma_{s, g} + \nu_{r, g})),
\end{equation}
where $\sigma$ is the logistic function. There are four latent variables with the following interpretation: $b_g$ represents the bias of exams in subgroup $g$, $\mu^{(n)}$ is the bias of exam $n$, $\gamma_{s, g}$ is the effect that low-pass filtering with severity $s$ has on exams in subgroup $g$, and $\nu_{r, g}$ is the idiosyncrasy of radiologist (DNN) $r$ on exams in subgroup $g$. See Figure~\ref{fig:pgm} for a graphical representation of our model. We considered several parameterizations of varying complexity, and selected the one with the maximum marginal likelihood. See the Methods section for details regarding our probabilistic model.

\paragraph{Comparing humans and machines with respect to their sensitivity to the removal of clinically meaningful information.}

\begin{figure}
    \centering
    \includegraphics[width=\textwidth]{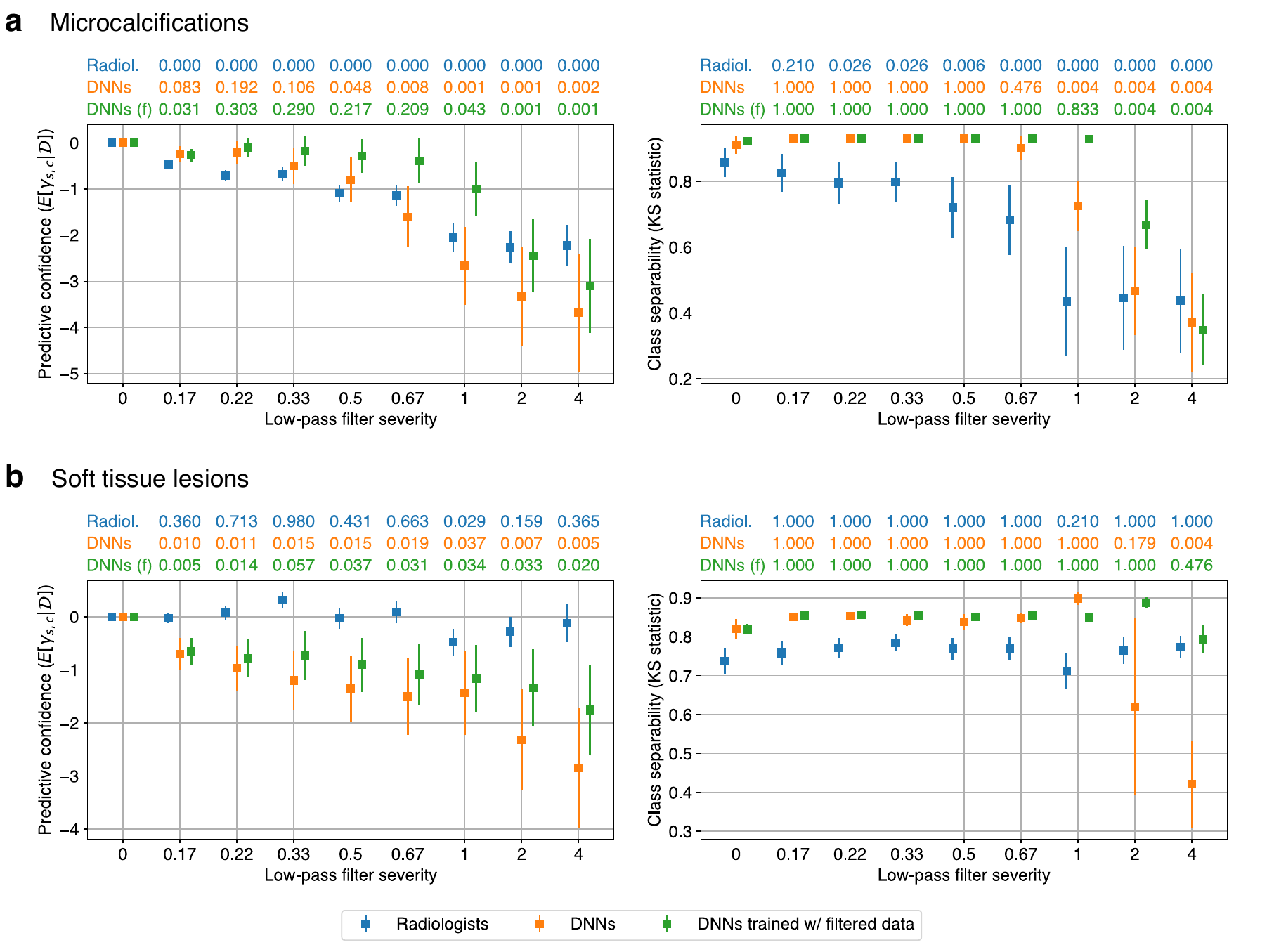}
    \caption{\textbf{Comparing human and machines with respect to their sensitivity to the removal of clinically meaningful information.} The left subfigures represent the effect on predictive confidence, measured as the posterior expectation of $\gamma_{s, g}$ for severity $s$ and subgroup $g$. The values at the top of each subfigure represent the probability that the predictive confidence for each severity is greater than zero. Smaller values for a given severity indicate a more significant downward effect on predictive confidence. The right subfigures correspond to the effect on class separability, quantified by the two-sample Kolmogorov-Smirnov (KS) statistic between the predictions for the positive and negative classes. The values at the top of each subfigure are the p-values of a one-tailed KS test between the KS statistics for a given severity and severity zero. Smaller values indicate a more significant downward effect on class separability for that severity. \textbf{(a)} For microcalcifications, low-pass filtering degrades predictive confidence and class separability for both radiologists and DNNs. When DNNs are trained with filtered data, the effects on predictive confidence and class separability are reduced, but not significantly. \textbf{(b)} For soft tissue lesions, filtering degrades predictive confidence and class separability for DNNs, but has no effect on radiologists. When DNNs are trained with filtered data, the effect on predictive confidence is reduced, and DNN-derived class separability becomes invariant to filtering.}
    \label{fig:gmic}
\end{figure}

Using the probabilistic model, we compared how low-pass filtering affects the predictions of radiologists and DNNs, separately analyzing microcalcifications and soft tissue lesions. We performed each comparison with respect to two metrics: predictive confidence and class separability. Since the latent variable $\gamma_{s, g}$ represents the effect of low-pass filtering on each prediction, we examined its posterior distribution in order to measure the effect on predictive confidence. We sampled values of $\hat{y}_{r, s}^{(n)}$ from the posterior predictive distribution in order to quantify how low-pass filtering affects class separability. We computed the Kolmogorov-Smirnov (KS) statistic between the sampled predictions for the positive and negative class. This represents the distance between the two distributions of predictions, or how separated the two classes are. Sampling from the posterior predictive distribution was necessary for radiologists, since we did not have a complete set of predictions from them. Although such sampling was not strictly necessary for DNNs given the full set of available predictions, we performed the same posterior sampling for DNNs in order to ensure a fair comparison.

Figure~\ref{fig:gmic}a presents the results for microcalcifications. We only consider DNNs that are trained with unperturbed data in this section. The results for training DNNs on low-pass filtered data are discussed in the next section. The left subfigure represents predictive confidence, as measured by the posterior expectation of $\gamma_{s, g}$. Since low-pass filtering removes the visual cues of malignant lesions, we hypothesized that it should decrease predictive confidence. In other words, we expected to see $\E[\gamma_{s, g} \mid \D] \leq 0$. Above the left subfigure, we report $P(\gamma_{s, g} > 0 \mid \D)$ in order to quantify how much the posterior distributions $P(\gamma_{s, g} \mid \D)$ align with this hypothesis. Small values indicate a significant negative effect on predictive confidence. We note that these values are not intended to be interpreted as the $p$-values of a statistical test. Instead, they quantify the degree to which each $\gamma_{s, g}$ is negative. We observe that for microcalcifications, low-pass filtering decreases the predictive confidence of both radiologists and DNNs. There is, however, an interesting difference in that for the range of most severe filters, the effect is constant for radiologists, while DNNs continue to become less confident.

The right subfigure of Figure~\ref{fig:gmic}a depicts the effect of low-pass filtering on class separability. This is quantified by the KS statistic between the predictions for the positive and negative class, where the positive class is restricted to malignant microcalcifications. Similar to our hypothesis that low-pass filtering decreases predictive confidence, we hypothesized that it should also reduce class separability. That is, we expected the KS statistics for severity $s > 0$ to be smaller than those for $s = 0$. This is because removing the visual cues for malignant lesions should make it more difficult to distinguish between malignant and nonmalignant cases. We tested this hypothesis using the one-tailed KS test between the KS statistics for $s = 0$ and $s > 0$. The $p$-values for this test are reported above the right subfigures, where small values mean that filtering significantly decreases class separability. We found that low-pass filtering decreases class separability for both radiologists and DNNs, but in substantially different ways. The radiologists' class separability steadily declines for the range of less severe filters, while it is constant for DNNs. Meanwhile, similar to what we observed for predictive confidence, the radiologists' class separability is constant for the range of most severe filters, while it continues to decline for DNNs. In summary, for microcalcifications, low-pass filtering decreases the predictive confidence and class separability for both radiologists and DNNs, but differences suggest that they use a different set of high frequency components.

Next, we compared how low-pass filtering affects the predictions of radiologists and DNNs on soft tissue lesions (Figure~\ref{fig:gmic}b). The results show that low-pass filtering degrades the predictive confidence and class separability of DNNs, while having almost no effect on radiologists. The invariance of radiologists' impressions to filtering implies that the high frequency components used by DNNs in this context are spurious, since there exist alternative features used by radiologists that are robust to low-pass filtering. The DNNs' reliance on such spurious features is a clear vulnerability, and must be addressed before they can be trusted to make clinical diagnoses. In the Discussion section, we describe a potential explanation for this phenomenon, as well as a plausible scenario in which this behavior would cause DNNs' performance to degrade.

\paragraph{Training DNNs with low-pass filtered data.}

We observed that low-pass filtering decreases the predictive confidence and class separability of DNNs for all lesion subgroups. However, since the DNNs only encountered low-pass filtering during testing, it is possible that this effect is solely due to the dataset shift between training and testing. We therefore repeated the previous experiments for DNNs, where the same filtering was applied during both training and testing. We then examined whether the effects of low-pass filtering on the DNNs' perception could be attributed to information loss rather than solely to dataset shift.

For microcalcifications (Figure~\ref{fig:gmic}a), training on filtered data slightly reduced the effect of low-pass filtering on predictive confidence and class separability, but the effect was still present, particularly for the most severe filters. This implies that the effect of filtering on microcalcifications can be attributed to information loss, and not solely to dataset shift. In other words, high frequency components in microcalcifications contain information that is important to the perception of DNNs.

Meanwhile, for soft tissue lesions (Figure~\ref{fig:gmic}b), training on low-pass-filtered data significantly reduces the effect on predictive confidence and class separability, even for severe filters. This suggests that the effect of low-pass filtering on soft tissue lesions can primarily be attributed to dataset shift rather than information loss. In fact, DNNs trained with low-pass-filtered data maintain a similar level of class separability compared to networks trained on the original data. This confirms what we observed for radiologists, which is that high frequency components in soft tissue lesions are largely dispensable, and that more robust features exist.

\paragraph{Annotation reader study.}

Our results thus far show that radiologists and DNNs use a different set of high frequency components in both microcalcifications and soft tissue lesions. Since this analysis has purely been in the frequency domain, we extend our comparison to the spatial domain by examining the degree to which radiologists and DNNs agree on the most suspicious regions of an image. Towards this end, we conducted a reader study in which seven radiologists annotated up to three regions of interest (ROIs) containing the most suspicious features of each image. 120 exams were used in this study, which is a subset of the 720 exams in the perturbation reader study. See the Methods section for details regarding this reader study. We then applied low-pass filtering to the interior and exterior of the ROIs, as well as to the entire image. Examples of the annotation and the low-pass filtering schemes are shown in Figure~\ref{fig:framework} (g--j). We made predictions using DNNs trained with the original data in order to understand the relationship between the high frequency components utilized by DNNs, and the regions of mammograms that are most suspicious to the radiologists.

\paragraph{Comparing humans and machines with respect to the regions of an image deemed most suspicious.}

\begin{figure}
    \centering
    \includegraphics[width=\textwidth]{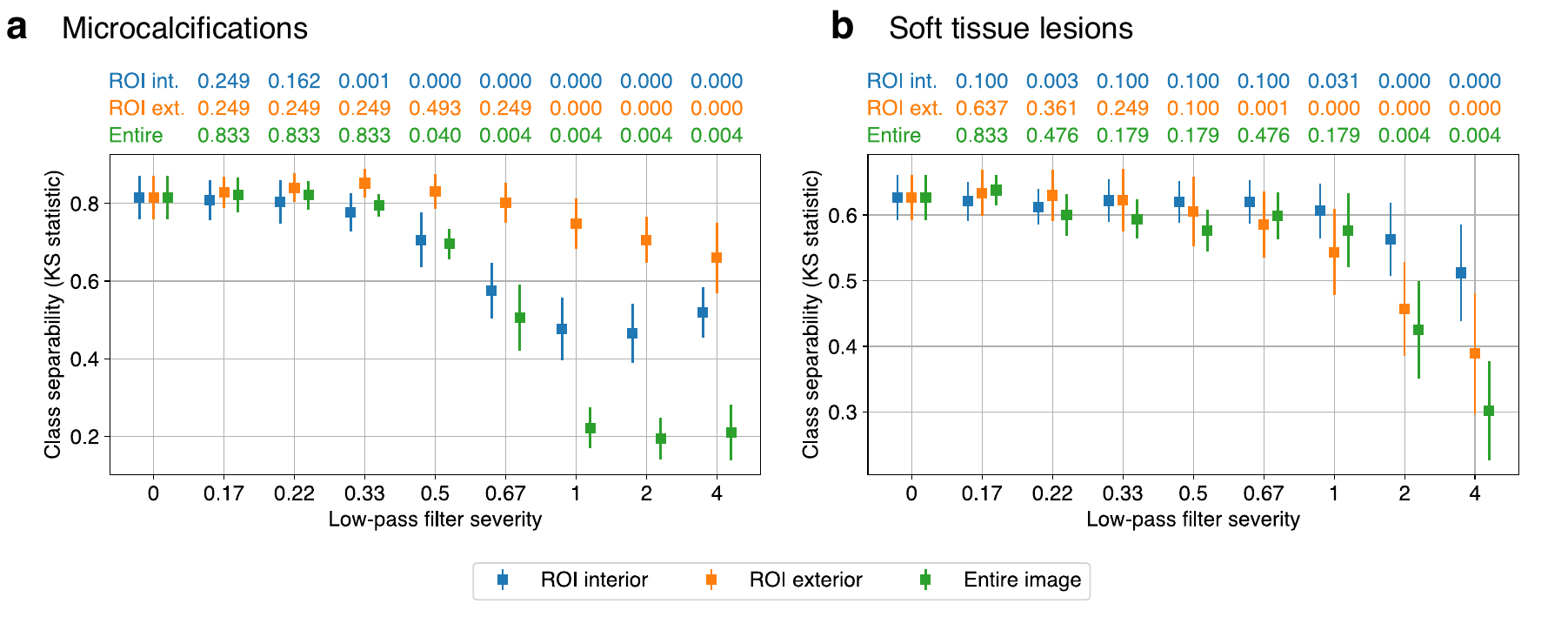}
    \caption{\textbf{Comparing humans and machines with respect to the regions of an image deemed most suspicious.} The performance of DNNs trained on unfiltered images was evaluated on images with selective perturbations in regions of interest (ROIs) identified as suspicious by human radiologists. \textbf{(a)} For microcalcifications, filtering the ROI interior decreases predictive confidence, but not as much as filtering the entire image. Filtering the ROI exterior decreases predictive confidence as well, meaning that DNNs utilize high frequency components in both the interior and the exterior of the ROIs, whereas humans focus more selectively on those ROIs. \textbf{(b)} For soft tissue lesions, filtering the ROI interior has very little effect on class separability. Meanwhile, filtering the ROI exterior has a similar effect to filtering the entire image. This implies that the high frequency components used by DNNs in these lesion subgroups are not localized in the areas that radiologists consider suspicious.}
    \label{fig:gmic_annotation_reader_study}
\end{figure}

We began by comparing the effect of the three low-pass filtering schemes on the DNNs' predictions for microcalcifications (Figure~\ref{fig:gmic_annotation_reader_study}a). We observed that filtering the ROI interior has a similar effect to filtering the entire image for mild filter severities. This suggests that for the frequencies in question, DNNs primarily rely on the same regions that radiologists consider suspicious. Meanwhile, class separability for the two ROI-based filtering schemes diverge significantly for high severities: filtering the ROI interior ceases to further decrease class separability at some threshold filter severity, whereas exterior filtering continues to degrade class separability beyond this threshold. The implication is that a range of high frequency components utilized by DNNs exist in the exterior of the ROIs deemed most important by human radioloigsts.

For soft tissue lesions (Figure~\ref{fig:gmic_annotation_reader_study}b), filtering the ROI interior decreases class separability, but to a lesser degree compared to filtering the entire image. This means that DNNs do utilize high frequency components in regions that radiologists find suspicious, but only to a limited degree. Meanwhile, filtering the ROI exterior has a similar effect on class separability as filtering the entire image. These observations suggest that the high frequency components that DNNs use for soft tissue lesions may be scattered across the image, rather than being localized in the areas that radiologists consider suspicious.

\section*{Discussion}

Arguably, the most significant difference we observed between human and machine perception in our case study was that for soft tissue lesions, DNNs use spurious high frequency components that are ignored by radiologists. Interestingly, we showed that by training DNNs on low-pass-filtered data, the difference narrows significantly in terms of predictive confidence, and nearly disappears for class separability. Here, we present an explanation for this phenomenon, as well as a realistic scenario in which it is harmful. We hypothesize that DNNs rely heavily on the borders of soft tissue lesions, while radiologists make more use of the radiodense or bright areas within the borders. Low-pass filtering removes the borders without affecting the interior brightness, which can explain why DNNs are severely affected while radiologists are invariant. If our hypothesis is true, this is potentially harmful since the borders of soft tissue lesions are often obscured in patients with high breast fibroglandular tissue density~\cite{Bae2014BreastDensity}. In other words, DNNs that rely on borders can be compromised by a dataset shift towards patients with breasts that are more dense than those in the training set. As for the effect of training on low-pass-filtered data, our interpretation is that DNNs are biased towards learning high frequency components, and this causes them to associate spurious high frequency patterns with malignant soft tissue lesions. When these spurious patterns are absent from the data, DNNs are able to learn alternative features with little effect on their class separability. Based on our hypothesis regarding the borders of soft tissue lesions and their bright interiors, we suspect that by training on low-pass-filtered data, DNNs are able to become more like radiologists by using the radiodense interiors of lesions rather than relying excessively on the borders. Training on low-pass-filtered data can therefore be interpreted as inducing a prior to favor low-frequency features.

\begin{figure}
    \centering
    \includegraphics[width=\textwidth]{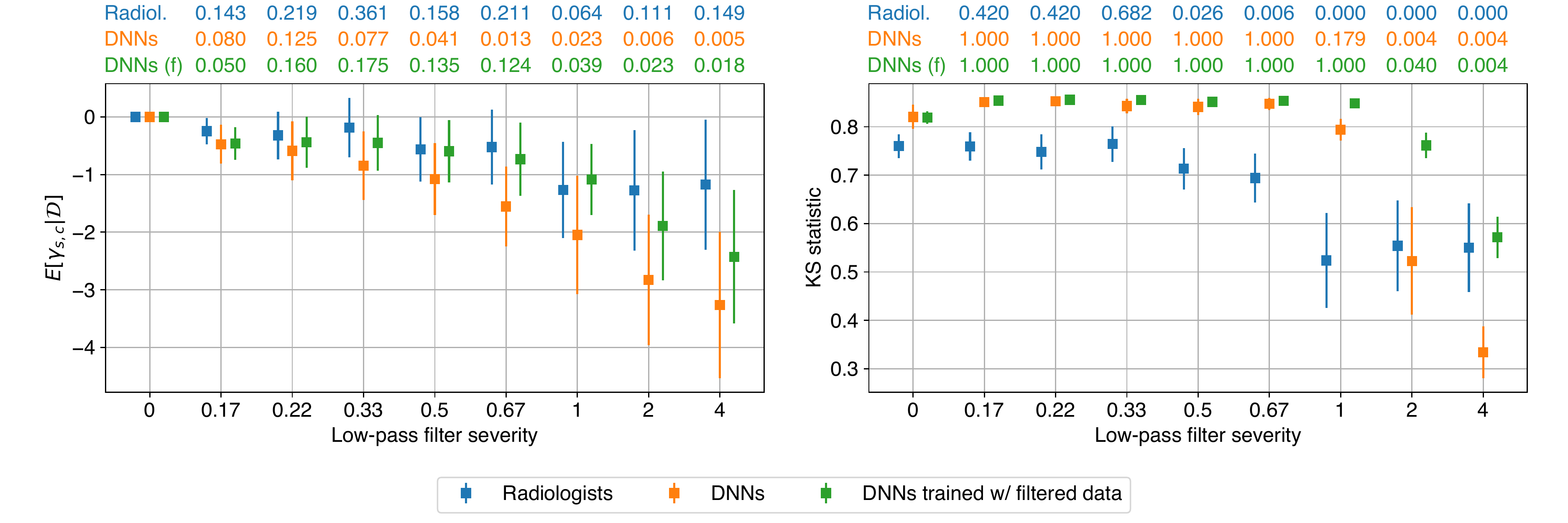}
    \caption{\textbf{Simpson's paradox leads to incorrect conclusions.} If we merged microcalcifications and soft tissue lesions into a single subgroup, we would incorrectly conclude that radiologists and DNNs exhibit similar perturbation robustness both for predictive confidence (left) and for class separability (right). This highlights the importance of performing subgroup analysis when comparing human and machine perception.}
    \label{fig:simpsons_paradox}
\end{figure}

Our framework draws inspiration from perturbation robustness studies in the adjacent domain of natural images, where there is also an ongoing crisis regarding the trustworthiness of machine perception. One key innovation in our work is to incorporate subgroup analysis to draw precise conclusions regarding perception. This is critically important for establishing trust in DNNs in the medical domain, where incorrect conclusions can directly harm patients' physical and mental well-being. Not accounting for subgroups can be very dangerous, as it can lead to drawing erroneous conclusions due to Simpson's paradox. In our case study, our conclusions would change significantly if we treated microcalcifications and soft tissue lesions as a single subgroup. As shown in Figure~\ref{fig:simpsons_paradox}, we would incorrectly conclude that radiologists and DNNs have comparable perturbation robustness in terms of both predictive confidence and class separability, thus artificially inflating the similarity between human and machine perception.

The identification of subgroups with clinically meaningful differences is a crucially important component of our framework, as it strongly influences the conclusions. It requires domain knowledge, and is not as simple as enumerating all possible subgroups. The reason is that, due to the rarity of some subgroups, there is a balance to strike between the number of specified subgroups and the amount of available data. In our case study, we combined architectural distortions, asymmetries, and masses into the soft tissue lesion subgroup because there are only 30 cases of malignant soft tissue lesions in our reader study dataset. By doing this, we addressed data scarcity while accounting for the fact that soft tissue lesions as a whole are much less localized than microcalcifications, and thus require a significantly different diagnostic approach.

The choice of input perturbation is another key consideration in our framework. For the purpose of understanding perception, it is more important for the removed information to be clearly characterizable than for the perturbation to be clinically realistic. For example, analyzing robustness in cross-institutional settings is clinically realistic, but it does not allow us to draw precise conclusions regarding perception, since it is unclear what information may have changed between institutions. Having said that, if the perturbation removes clinically relevant information and is additionally clinically realistic, this is beneficial because it allows us to reason about robustness in a plausible scenario. Our choice of Gaussian low-pass filtering is clinically relevant, as another type of low-pass filtering called motion blurring does occur in practice. Mammograms can be blurred by motion caused by patients or imaging devices~\cite{Choi2014MammographicArtifacts}, and \cite{Abdullah2017MotionBlur} demonstrated that it can degrade the ability of radiologists to detect malignant lesions. While there exist differences between Gaussian low-pass filtering and motion blurring, we expect that robustness to the former will translate to the latter. This is because DNNs have been shown to exhibit similar robustness characteristics to various types of blurring~\cite{Hendrycks2019RobustnessBenchmark}. We noticed that when comparing the class separability between radiologists and DNNs trained with low-pass-filtered data (Figure~\ref{fig:gmic}), DNNs are more robust to low-pass filtering for microcalcifications, while both are largely invariant for soft tissue lesions. This may be a significant advantage of DNNs in clinical practice.

In summary, we proposed a framework for establishing trust in machine perception for medical diagnosis, which we expect to be applicable to a variety of clinical tasks and imaging technologies. We demonstrated the efficacy of this framework with a case study in breast cancer screening. The framework uses subgroup-specific perturbation robustness to compare human and machine perception along two axes: their sensitivity to the removal of clinically meaningful information, and the regions of an image deemed most suspicious. In our case study, we revealed significant differences between radiologists and DNNs along both axes of comparison. We found that for microcalcifications, DNNs use a different set of high frequency components than radiologists, and some of these components lie outside the image regions that radiologists find most suspicious. Meanwhile, for soft tissue lesions, DNNs utilize spurious high frequency components that are ignored by radiologists. We also showed that we would have missed this stark divergence between radiologists and DNNs in soft tissue lesions if we failed to perform subgroup analysis. This is evidence that future studies comparing human and machine perception in other medical domains should separately analyze subgroups with clinically meaningful differences. By utilizing appropriate subgroup analysis driven by clinical domain knowledge, we can draw precise conclusions regarding machine perception, and potentially accelerate the widespread adoption of DNNs in clinical practice.

\printbibliography[segment=\therefsegment, check=onlynew]

\pagebreak

\section*{Methods}

\newrefsegment

\paragraph{DNN training methodology.}
We conducted our DNN experiments using two architectures: the Deep Multi-View Classifer~\cite{Wu2019Mammography}, and the Globally-Aware Multiple Instance Classifier~\cite{Shen2019GMIC,Shen2020GMIC}. With both architectures, we trained an ensemble of five models. A subset of each model's weights was initialized using weights pretrained on the BI-RADS label optimization task~\cite{Geras2017BIRADS}, while the remaining weights were randomly initialized. For each architecture, we adopted the same training methodology used by the corresponding authors.

\paragraph{Probability calibration.}
We applied Dirichlet calibration~\cite{Kull2019DirichletCalibration} to the predictions of DNNs used in our probabilistic modeling. This amounts to using logistic regression to fit the log predictions to the targets. We trained the logistic regression model using the validation set, and applied it to the log predictions on the test set to obtain the predictions used in our analysis. We used L2 regularization when fitting the logistic regression model, and tuned the regularization hyperparameter via an internal 10-fold cross-validation where we further split the validation set into ``training'' and ``validation'' sets. In the cross-validation, we minimized the classwise expected calibration error~\cite{Kull2019DirichletCalibration}.

\paragraph{Gaussian low-pass filtering.}
Low-pass filtering is a method for removing information from images that allows us to interpolate between the original image and, in the most extreme case, an image where every pixel has the value of the mean pixel value of the original image. We experimented with nine filter severities selected to span a large range of the frequency spectrum. We implemented the Gaussian low-pass filter by first applying the shifted two-dimensional discrete Fourier transform to transform images to the frequency domain. The images were multiplied element-wise by a mask with values in $[0, 1]$. The values of this mask are given by the Gaussian function
\begin{equation}
    M(u, v) = \exp\left(\frac{-D^2(u, v)}{2D^2_0}\right),
\end{equation}
where $u$ and $v$ are horizontal and vertical coordinates, $D(u, v)$ is the Euclidian distance from the origin, and $D_0$ is the cutoff frequency. $D_0$ represents the severity of the filter, where frequencies are reduced to $0.607$ of their original values when $D(u, v) = D_0$. Since the mammograms in our dataset vary in terms of spatial resolution as well as the physical length represented by each pixel, we expressed the filter severity $D_0$ in terms of a normalized unit of cycles per millimeter on the breast. Let $\alpha = \min(H, W)$ where $H$ and $W$ are the height and width of the image, and let $\beta$ denote the physical length in millimeters represented by each pixel. Then we can convert cycles per millimeter $D_0$ to cycles per frame length of the image $D_0^{\text{img}}$ using
\begin{equation}
    D_0^{\text{img}} = D_0 \cdot \alpha \cdot \beta.
\end{equation}

\paragraph{Perturbation reader study.}
In order to compare humans and machines with respect to their sensitivity to the removal of clinically meaningful information, we conducted a reader study in which ten radiologists read 720 exams selected from the test set. While all radiologists read the same set of exams, each exam was low-pass filtered with a different severity for each radiologist. Except for the low-pass filtering, the setup of this reader study is identical to that of~\cite{Wu2019Mammography}. Each exam consists of at least four images, with one or more images for each of the four views of mammography: L-CC, R-CC, L-MLO, and R-MLO. All images in the exam were concatenated into a single image such that the right breast faces left and is presented on the left, and the left breast faces right and is displayed on the right. Additionally, the craniocaudal (CC) views are on the top row, while the mediolateral oblique (MLO) views are on the bottom row. An example of this is shown in Extended Data Figure~\ref{fig:mammogram}. Among the 1,440 breasts, 62 are malignant, 356 are benign, and the remaining 1,022 are nonbiopsied. Among the malignant breasts, there are 26 microcalcifications, 21 masses, 12 asymmetries, and 4 architectural distortions, while in the benign breasts, the corresponding counts are: 102, 87, 36, and 6. For each exam, radiologists make a binary prediction for each breast, indicating their diagnosis of malignancy.

\paragraph{Probabilistic modeling.}
We modeled the radiologists' and DNNs' binary malignancy predictions with the Bernoulli distribution
\begin{equation}
    \hat{y}_{r, s}^{(n)} \sim \bernoulli(p_{r, s}^{(n)}),
\end{equation}
where $n \in \{1, 2, \dotsc, 1440\}$ indexes the breast, $r \in \{1, 2, \dotsc, 10\}$ the reader, and $s \in \{1, 2, \dotsc, 9\}$ the low-pass filter severity. Each distribution's parameter $p_{r, s}^{(n)}$ is a function of four latent variables
\begin{equation*}
    p_{r, s}^{(n)} = \sigma(b_g + \mu^{(n)} + \gamma_{s, g} + \nu_{r, g}),
\end{equation*}
where $c \in \{1, \dotsc, 5\}$ indexes the subgroup of the lesion. We included the following subgroups: unambiguous microcalcifications, unambiguous soft tissue lesions, ambiguous microcalcifications and soft tissue lesions, mammographically occult, and nonbiopsied. We considered these five subgroups in order to make use of all of our data, but only used the first two in our analysis. We assigned the generic weakly informative prior $\N(0, 1)$ to each latent variable. We chose the Bernoulli distribution because it has a single parameter, and thus makes the latent variables interpretable. Additionally, radiologists are accustomed to making discrete predictions in clinical practice. The posterior distribution of the latent variables is given by
\begin{equation*}
    p(\boldsymbol{b}, \boldsymbol{\mu}, \boldsymbol{\gamma}, \boldsymbol{\nu} \mid \boldsymbol{\hat{y}}) = \frac{p(\boldsymbol{\hat{y}} \mid \boldsymbol{b}, \boldsymbol{\mu}, \boldsymbol{\gamma}, \boldsymbol{\nu})}{p\left(\boldsymbol{\hat{y}}\right)}.
\end{equation*}
The exact computation of the posterior is intractable, since the marginal likelihood $p(\boldsymbol{\hat{y}})$ involves a four-dimensional integral. We therefore applied automatic differentiation variational inference (ADVI)~\cite{Kucukelbir2015ADVI} in order to approximate the posterior. ADVI, and variational inference in general, optimizes over a class of tractable distributions in order to find the closest match to the posterior. For our choice of tractable distributions, we used the mean-field approximation, meaning that we optimized over multivariate Gaussians with diagonal covariance matrices.

In practice, while radiologists made binary predictions, DNNs made continuous predictions in $[0, 1]$ that we then calibrated. Despite the DNN predictions not being binary, we used equivalent procedures to specify the probabilistic model for radiologists and DNNs. To see how, let $\hat{y}^{(n)} \in \{0, 1\}$ denote a DNN's unobserved binary prediction for case $n$, and let $\hat{z}^{(n)} \in [0, 1]$ denote its observed real-valued prediction for the same case. In order to obtain $\hat{y}^{(n)} \in \{0, 1\}$, we could treat it as a random variable $\hat{y}^{(n)} \sim \bernoulli(\hat{z}^{(n)})$ and obtain values for it through sampling. We instead used $\hat{z}^{(n)}$ directly, specifying the log joint density as
\begin{equation*}
	\begin{split}
	\log p(\boldsymbol{\hat{y}} \mid \theta) &= \sum_{n=1}^{N} \log p(\hat{y}^{(n)} \mid \theta^{(n)})\\
	&\approx \sum_{n=1}^{N} \E_{\hat{y}^{(n)}}[\log p(\hat{y}^{(n)} \mid \theta^{(n)})]\\
	&= \sum_{n=1}^{N} \E_{\hat{y}^{(n)}}\left[\hat{y}^{(n)} \log(\theta^{(n)}) +  (1 - \hat{y}^{(n)}) \log(1 - \theta^{(n)})\right]\\
	&= \sum_{n=1}^{N} \Pr(\hat{y}^{(n)} = 1) \log(\theta^{(n)}) + \Pr(\hat{y}^{(n)} = 0) \log(1 - \theta^{(n)})\\
	&= \sum_{n=1}^{N} \hat{z}^{(n)} \log(\theta^{(n)}) + (1 - \hat{z}^{(n)}) \log(1 - \theta^{(n)}).
	\end{split}
\end{equation*}

\paragraph{Annotation reader study.}
In order to compare humans and machines with respect to the regions of an image deemed most suspicious, we conducted a reader study in which seven radiologists read the same set of 120 unperturbed exams. The exams in this study were a subset of the 720 exams from the perturbation reader study, and also included all malignant exams from the test set. This study had two stages. In the first stage, the radiologists were presented with all views of the mammogram, and they made a malignancy diagnosis for each breast. This stage was identical to the reader study in \cite{Wu2019Mammography}. In the second stage, for breasts that were diagnosed as malignant, the radiologists annotated up to three ROIs around the regions they found most suspicious. The radiologists annotated each view individually, and the limit of three ROIs applied separately to each view. For exams that contained multiple images per view, the radiologists annotated the image where the malignancy was most visible. The radiologists annotated the images using Paintbrush on MacOS, or Microsoft Paint on Windows. In order to constrain the maximum area that is annotated for each image, we included a $250{\times}250$ pixel blue ROI template in the bottom corner of each image to serve as a reference. The radiologists then drew up to three red ROIs such that each box approximately matched the dimensions of the reference blue ROI template.

\section*{Data availability}

The radiologist and DNN predictions used in our analysis are available at \url{https://github.com/nyukat/perception_comparison} under the GNU AGPLv3 license.

\section*{Code availability}

The code used in this research is available at \url{https://github.com/nyukat/perception_comparison} under the GNU AGPLv3 license. This, combined with the predictions which we also open-sourced, makes our results fully reproducible. We used several open-source libraries to conduct our experiments. The DNN experiments were performed using PyTorch~\cite{Paszke2015PyTorch}, and the probabilistic modeling was done with PyStan (\url{https://github.com/stan-dev/pystan}), the Python interface to Stan~\cite{Carpenter2017Stan}. The code for the DNNs used in our experiments is also open-source, where GMIC is available at \url{https://github.com/nyukat/GMIC}, and DMV at \url{https://github.com/nyukat/breast_cancer_classifier}.

\printbibliography[segment=\therefsegment, check=onlynew]

\section*{Acknowledgements}

The authors would like to thank Mario Videna, Abdul Khaja and Michael Costantino for supporting our computing environment and Eric K. Oermann for helpful comments on the draft of this paper. We also gratefully acknowledge the support of NVIDIA Corporation, which donated some of the GPUs used in this research. This work was supported in part by grants from the National Institutes of Health (P41EB017183 and R21CA225175) and the National Science Foundation (HDR-1922658).

\section*{Author contributions}

SJ conceived the initial idea and designed the first set of experiments. TM, SJ, WO, KC and KJG designed the final version of the experiments. TM and WO conducted the experiments with neural networks. TM and WO preprocessed the screening mammograms. TM, SJ, LM and LH conducted the reader study. CeC, NS, RE, DQ, ChC, LD, EK, AK, JL, JP, KP, BR and HT collected the data. TM, SJ and WO conducted literature search. TM, KC and KJG designed the probabilistic model. TM performed the probabilistic inference. LM, LH and BR analyzed the results from a clinical perspective. DS contributed to framing the paper. SJ, KC and KJG supervised the execution of all elements of the project. DS, LM, LH, KC and KJG acquired funding for this research. All authors provided critical feedback and helped shape the manuscript.

\section*{Competing interests}

The authors declare no competing interests.

\pagebreak

\section*{Extended data}

\begin{figure}
    \centering
    \includegraphics[width=\textwidth]{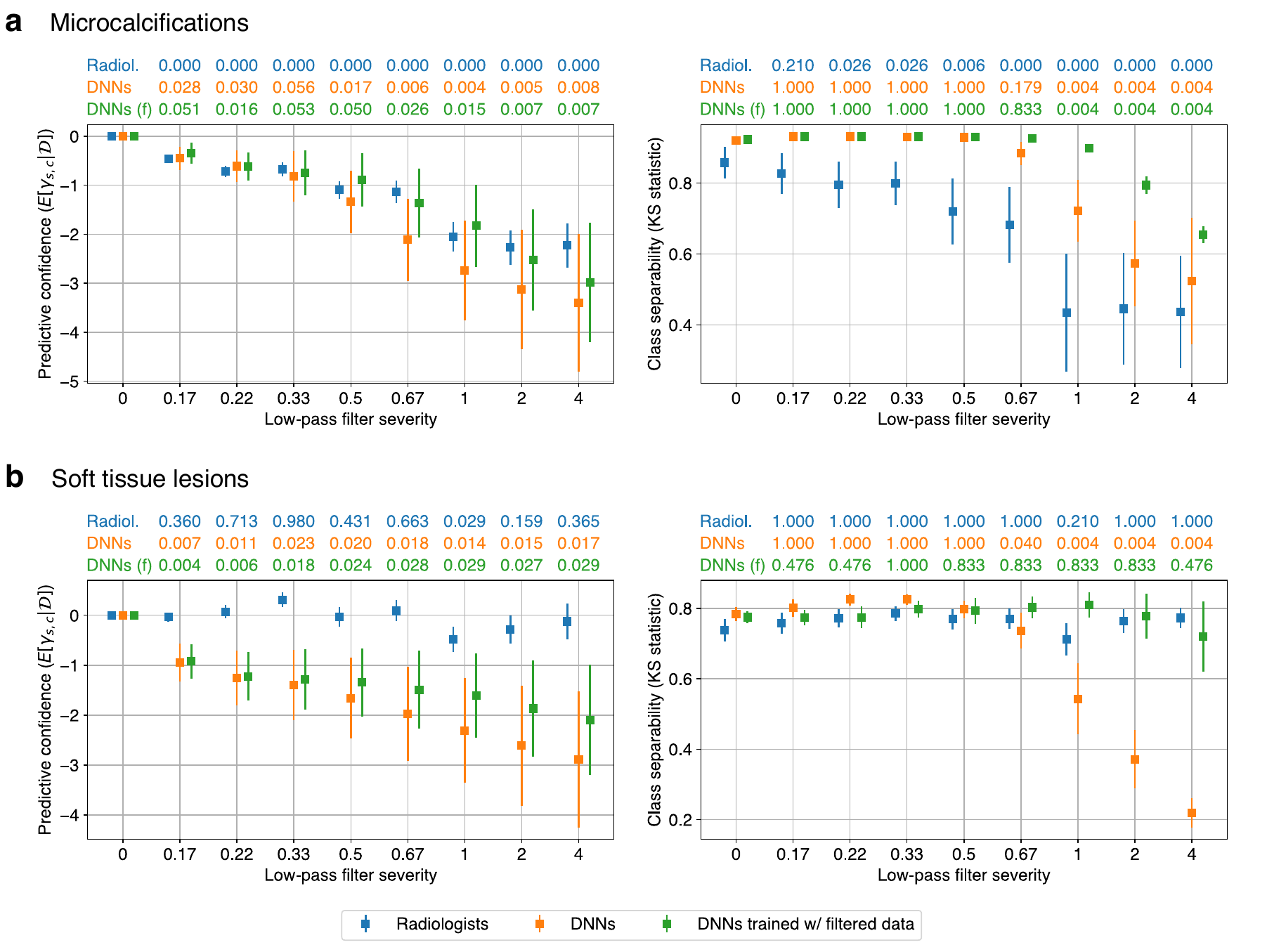}
    \caption{\textbf{Comparing humans and machines with respect to their sensitivity to the removal of clinically meaningful information.} These results are for the DMV architecture, and our conclusions are the same as with the GMIC architecture.}
    \label{fig:dmv}
\end{figure}

\begin{figure}
    \centering
    \includegraphics[width=\textwidth]{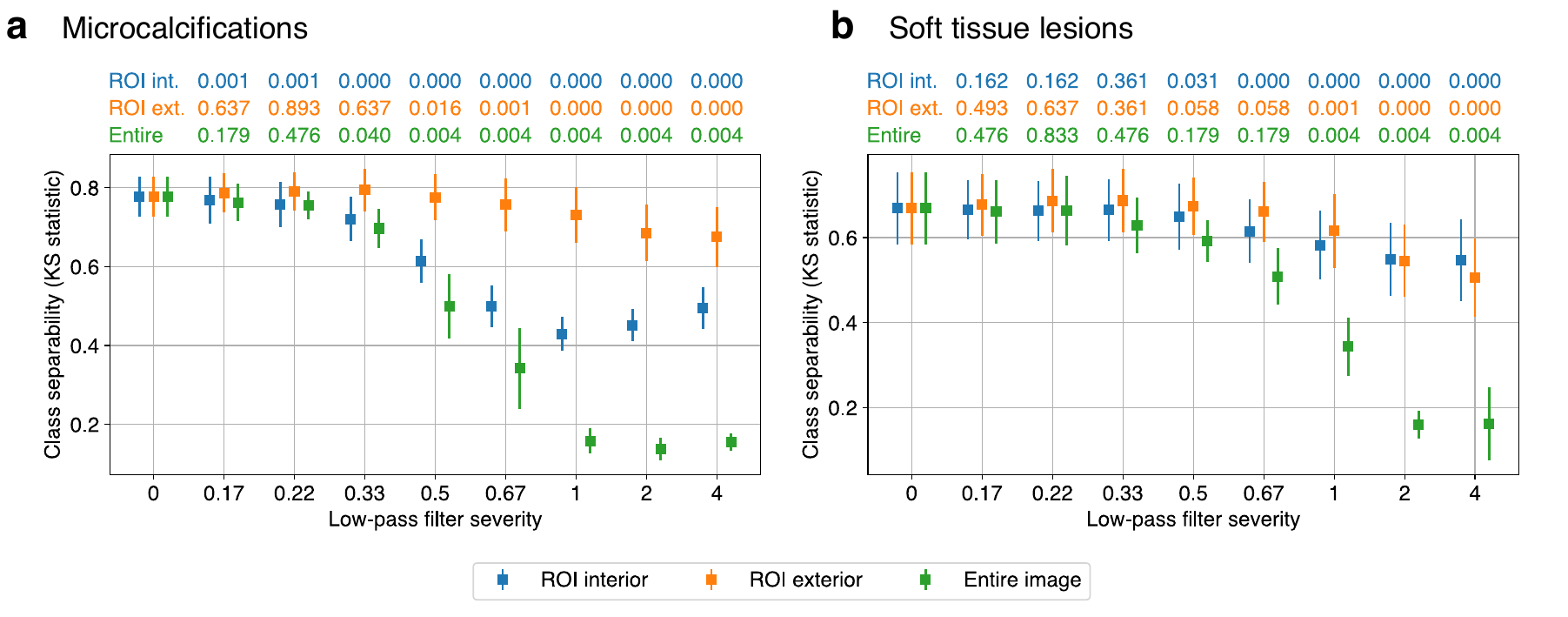}
    \caption{\textbf{Comparing humans and machines with respect to the regions of an image deemed most suspicious.} These results are for the DMV architecture, and our conclusions are the same as with the GMIC architecture.}
    \label{fig:dmv_annotation_reader_study}
\end{figure}

\end{document}